\documentclass[aip,jcp,preprint]{revtex4-1}

\usepackage{graphicx}
\usepackage[latin1]{inputenc}
\usepackage{braket}
\usepackage{units}
\usepackage{wrapfig}
\usepackage{graphicx}
\usepackage{amsmath}
\usepackage{booktabs}
\usepackage{rotating}
\usepackage{color}
\usepackage{amsfonts}
\usepackage{amssymb}
\usepackage{moreverb}
\usepackage{url}
\usepackage{epsf}
\usepackage{multirow}
\usepackage{subfigure}
\usepackage{morefloats}
\usepackage{threeparttable}
\usepackage{cancel}
\usepackage{bm}
\usepackage{multirow}

\begin{document}

\title{Selecting Boron Fullerenes by Cage-Doping Mechanisms}

\author{Paul Boulanger} 
\homepage{http://inac.cea.fr/L_Sim}
\affiliation{Laboratoire de simulation atomistique (L\_Sim), SP2M, INAC, CEA-UJF, Grenoble, F-38054, France} 

\author{Maxime Morini\`ere}
\affiliation{Laboratoire de simulation atomistique (L\_Sim), SP2M, INAC, CEA-UJF, Grenoble, F-38054, France}

\author{Luigi Genovese}
\affiliation{Laboratoire de simulation atomistique (L\_Sim), SP2M, INAC, CEA-UJF, Grenoble, F-38054, France}

\author{Pascal Pochet}
\email{pascal.pochet@cea.fr}
\affiliation{Laboratoire de simulation atomistique (L\_Sim), SP2M, INAC, CEA-UJF, Grenoble, F-38054, France}

\date{\today}

\begin{abstract}
So far, no boron fullerenes were synthesized: more compact  $sp^3$-bonded clusters are energetically
preferred. To circumvent this, metallic clusters have been suggested by Pochet \textit{et al.} [Phys. Rev. B 83, 081403(R) (2011)] as ``seeds'' for a possible
synthesis which would topologically protect the $sp^2$ sector of the configuration space. In this paper, we identify a
basic pentagonal unit which allows a balance between the release of strain and the self-doping rule. We formulate a
guiding principle for the stability of boron fullerenes, which takes the form of an isolated filled pentagon rule
(IFPR). The role of metallic clusters is then reexamined. It is shown that the interplay of the IFPR and the
seed-induced doping breaks polymorphism and its related problems:  it can effectively select between different isomers and
reduce the reactivity of the boron shells. The balance between self and exterior doping represents the best strategy for boron
buckyball synthesis. 
\end{abstract}


\maketitle

\section{Introduction}
Boron analogues \cite{doi:10.1021/ar0300266} to carbon $sp^2$ structures usually
take the form of compounds where an other element acts as an electron donor. In
these structures, the $\mathrm{B}^-$ anions behave like carbon atoms. This is
the case of $\mathrm{MgB}_2$ \cite{doi:10.1021/ja01634a089,MgB2_nature}, the boron analogue of graphite.
For less extended structures this purely exterior doping can be unfeasible and
impractical, e.g. giving 60 electrons to a boron buckyball is no simple
task. Fortunately,  simulations have shown that pure boron systems can partially
overcome the low electron count on their own through self-doping
\cite{PhysRevB.80.134113}. This mechanism explains the stability of small planar
$sp^2$ clusters \cite{nanosheet,nanosheet2,Boustani1997182}, the relative stability of the all
boron $\alpha$-sheet \cite{PhysRevLett.99.115501} and other structures
\cite{PhysRevB.74.035413,doi:10.1021/nl073295o,Zope2011193}. In contrast with carbon,
these all-boron structures were also expected to display polymorphism
\cite{doi:10.1021/nl3004754,doi:10.1021/nn302696v,PhysRevLett.106.225502} and
multi-center bonding \cite{PhysRevLett.99.115501,C1CP20439D} thus allowing new
behaviors for well-known structures.  On this basis, the $\mathrm{B}_{80}$
fullerene has attracted a lot of attention
\cite{PhysRevLett.98.166804,PhysRevB.79.161403,PhysRevB.78.201401,
PhysRevB.82.153409,PhysRevB.83.081403,li:074302}.

However, no boron fullerene were observed experimentally. Consequently, it was shown that other
less symmetric bulk-like precursors, dubbed core-shell structures, are
energetically preferred
\cite{PhysRevLett.100.165504,C002954H,doi:10.1021/jp1018873,
PhysRevLett.106.225502}. Self-doping is thus insufficient to guarantee $sp^2$
bonding in extended boron. Nevertheless,  it can be argued that synthesis of these systems
remains possible if both self and exterior doping are present. Endohedral metalloborofullerenes
\cite{wang:133102,doi:10.1021/jp9019848} were found to be energetically more
stable than the corresponding core-shell clusters \cite{PhysRevB.83.081403},
suggesting a possible synthesis pathway which is very similar to the one used in
the synthesis of endohedral carbon fullerenes.  The recent observation of boron
nanotubes inside Mg rich catalytic pores \cite{doi:10.1021/jp049301b} or on catalytic surfaces
\cite{B919260C} give even more weight to this scenario.

Yet, some problems would remain. The polymorphism of boron can also prove to be a problem
since there would be no energetically motivated driving force towards some
well-defined structure \cite{PhysRevLett.106.225502}. A successful synthesis
would lead to disordered structures which could negatively impact their
characterization and properties. Furthermore, polymorphism is also related to
high reactivity: endohedral boron fullerenes could still have the possibility of
forming outward $sp^3$ boron structures by merging with other boron clusters
during synthesis.



\begin{figure}
\begin{center}
\includegraphics[scale=0.3]{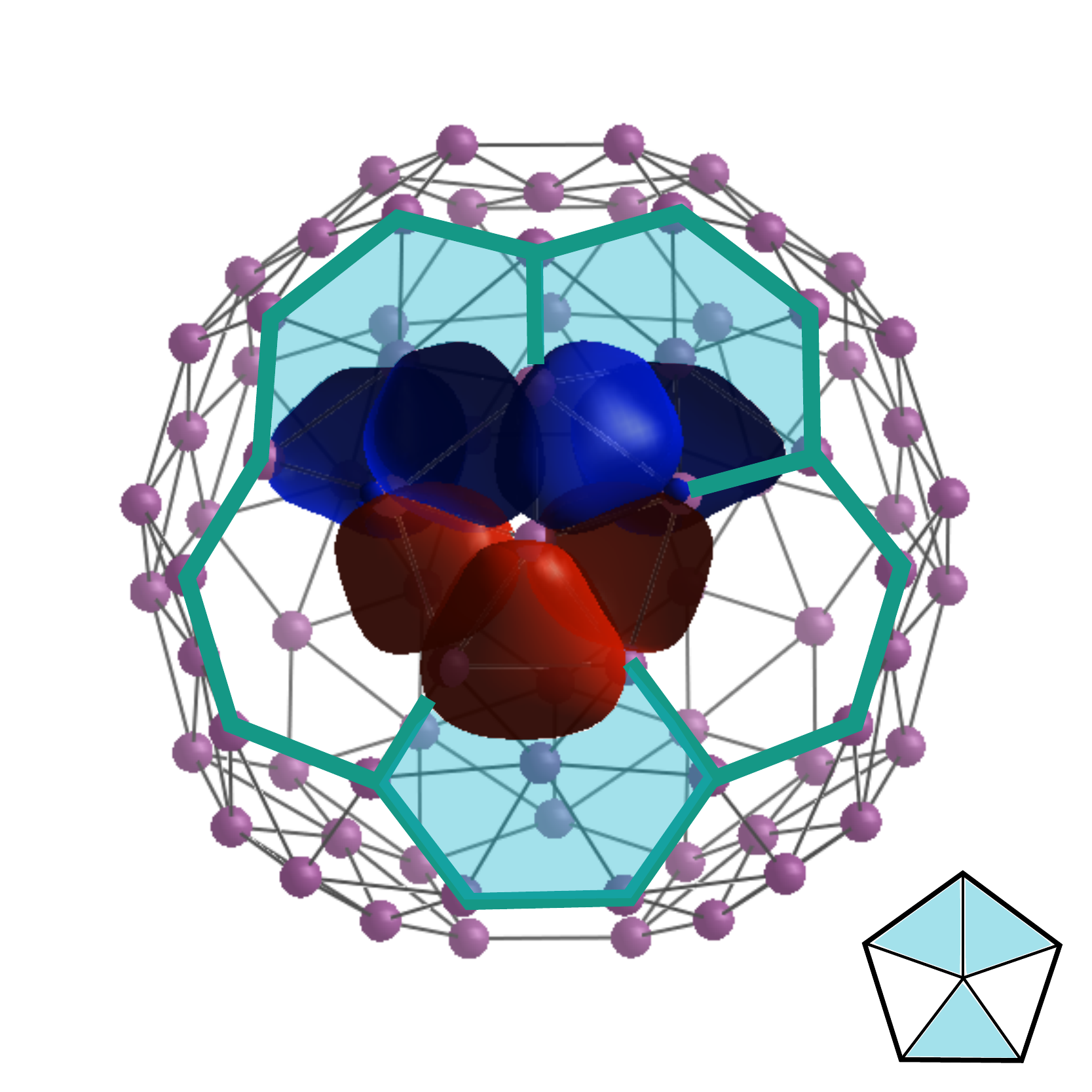}
\caption{Wannier functions representing the bonding pattern of a filled pentagon for the
$\mathrm{C}_{3}$ 17-3 isomer. The 3c-2e Wannier functions are represented in red
and the 4c-2e Wannier functions are in blue. The neighboring filled hexagons are shaded to
clarify the local environment. A schematic representation of this pentagonal
environment is shown at the bottom right. This coloring scheme is reused in Figure \ref{Fig:systems}.}
\label{Fig:pentagon_bonding}
\end{center}
\end{figure}

\begin{figure*}
\begin{center}
\includegraphics[scale=0.3]{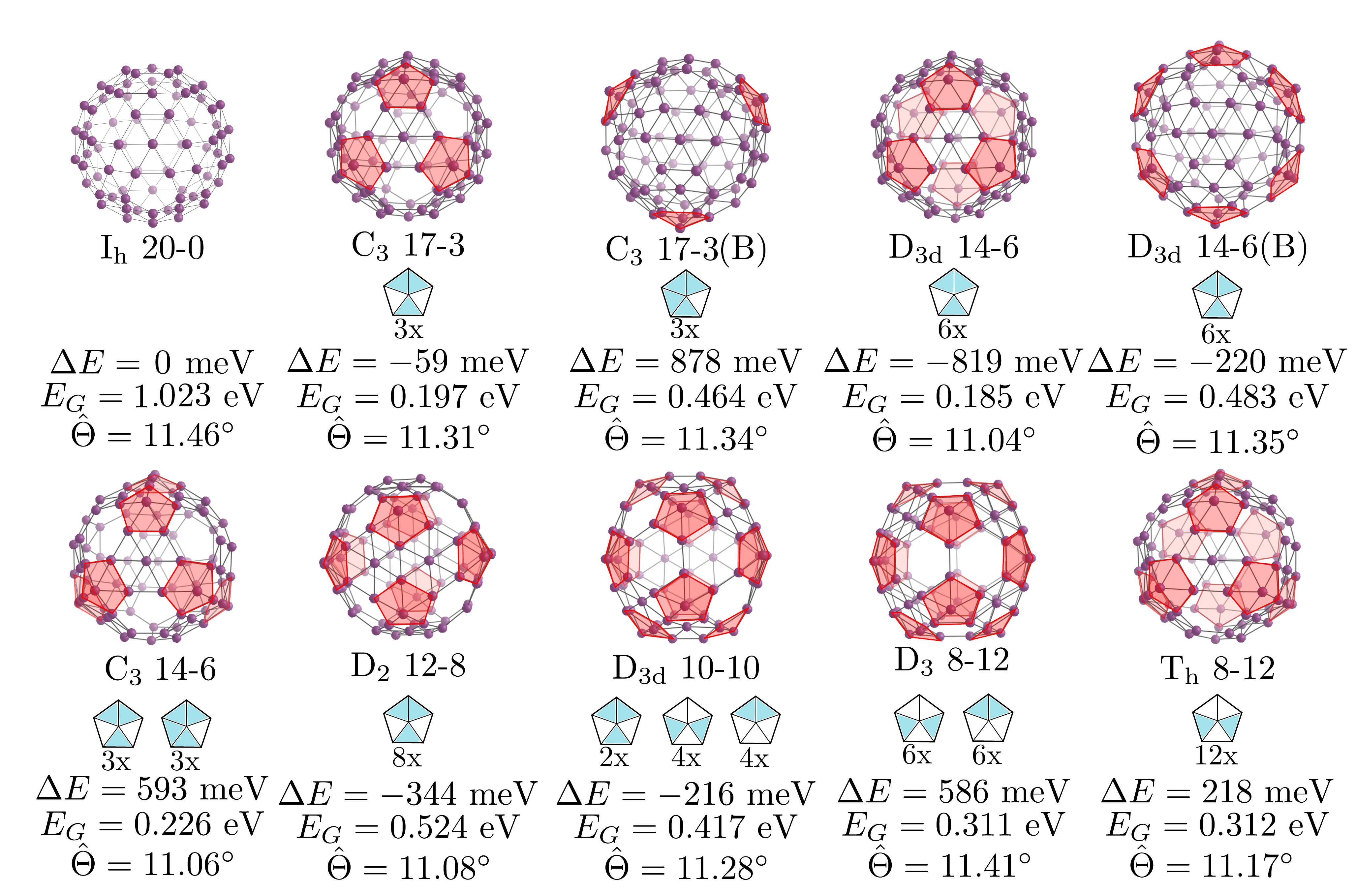}
\caption{The different $B_{80}$ isomers with increasing number
of filled pentagons. They were chosen to exemplify the role of the filled
pentagonal unit inside these structures. The different local environments of
the filled pentagons which are present in each isomer are sketched below them along with their frequency.
The colored sections of these sketches represent a neighboring filled hexagon. Also
displayed is the energy difference ( $\Delta E$ in meV) between the isomer and the
$\mathrm{I}_{\mathrm{h}}$ 20-0, its HOMO-LUMO gap ($\mathrm{E}_G$ in eV) and its average pyramidalization angle ($\hat{\Theta}$).}
\label{Fig:systems}
\end{center}
\end{figure*}
In our paper, we offer a critical reassessment of the role of doping in the boron fullerenes and show how it can
overcome the above mentioned problems. We first rationalize the role of self-doping \cite{PhysRevB.80.134113}
in the pure $\mathrm{B}_{80}$ borofullerene. The different $\mathrm{B}_{80}$ fullerene isomers are formed of a
$\mathrm{B}_{60}$ backbone with an additional 20 filling boron atoms, which can be placed in the center of hexagons or pentagons.
These boron atoms act as dopants. We show that filled pentagons contribute to lower the global strain of the fullerene, but a specific
type of local environment is necessary to satisfy the electron-counting issues arising from the self-doping rule for filled pentagons.
Additionally, it will be shown that delocalized  $\pi$ bond percolation (partial aromaticity) plays a stabilizing role in these isomers. All of these findings support
the fact that, for these open-shell clusters, the pentagonal pyramids represent the crucial building block for
the stability of the $\mathrm{B}_{80}$ fullerenes. This leads us to formulate a generalized isolated filled pentagon rule (IPFR). 
Having understood the consequences of the strain and of self-doping in pure borofullerenes,
we then focus on the role of exterior doping, i.e. charge transfer via endohedral centers (seeds) or anions. We show that
exterior doping not only further stabilizes these clusters but also diminishes their reactivities. Furthermore, the
overall polymorphism can be broken by the interaction with a seed. Hence, endohedral metalloborofullerenes possess
all the required properties to overcome the known hurdles inhibiting borofullerene synthesis.

\section{Methods}
We use the Kohn-Sham approach to density-functional theory (DFT) as implemented
in the BigDFT code \cite{genovese:014109}, which uses a systematic real-space
wavelet basis. We considered the basis converged when an accuracy of 0.5
meV/atom was reached for the total energy and 1 meV/\AA ~for the forces. This
corresponds to a uniform wavelet grid with a spacing of 0.4 bohrs with an extent
of 11.5 bohrs for the coarse grid and 2.9 bohrs for the fine grid. The
structural relaxations used the FIRE algorithm \cite{PhysRevLett.97.170201} and were stopped when the forces
were below 5 meV/\AA. In addition HGH pseudopotentials \cite{PhysRevB.58.3641} in the Krack variant
\cite{springerlink:10.1007/s00214-005-0655-y} were used with isolated boundary conditions.
Hence, no supercell approximation was needed. 

Our calculations used the GGA-PBE \cite{PhysRevLett.77.3865} exchange-correlation
functional, which was shown to give the same energy organization as coupled cluster
methods for small boron clusters (in contrast to other hybrid functionals) \cite{li:074302}.
The GGA-PBE was also shown to yield more accurate results than LDA in these systems,
even if some deviation from diffusion quantum Monte Carlo was seen \cite{PhysRevB.79.245401}.

The Wannier functions were extracted from the Kohn-Sham wavefunctions using the Wannier90 
code \cite{Mostofi2008685}. They were considered converged when the total spread varied by
less then $10^{-12} \AA^2$. An initial projection on $sp^2$ functions located on the atoms
and with the $z$ axis pointing along the radial direction was needed. 
In order to quantify the degree of strain inherent to each isomer, we have used the Pi Orbital Axis Vector (POAV2)
method \cite{doi:10.1021/ar00150a005} developed for carbon fullerenes. The POAV2 analysis 
of the structure was also done with a code developed for this purpose. The pyramidalization angle is defined as the mean
deviation from perfect orthogonality between the $\sigma$ and $\pi$ orbitals. The multi-center bonding 
revealed by the Wannier analysis (or other local orbitals analysis \cite{C1CP20439D}) coupled to the high
coordination number of boron atoms (5 or 6) adds some difficulties in the POAV analysis of these
systems. Luckily, a study of the Wannier functions revealed that the filling atoms do not
possess local orbitals with $\pi_z$ character. They thus literally behave as doping atoms which 
contributes $\sigma$ electrons in three 3c-2c bonds. Hence, there is no point in evaluating 
there $\pi$ misalignment. On the other hand, the backbone atoms do possess $\pi_z$ character and
the POAV2 analysis was done only for these atoms. The still relatively high coordination of these
atoms (5) did not pose a problem because the 3c-2e Wannier centers were located on the plane defined
by the corresponding three atoms and thus using any of these atoms yields the same pyramidalization angle.
Thus, for simplicity, only the backbone atoms were used for this analysis which reduced the effective coordination
to three.  

In order to be concise, we do not discuss the pseudo-Jahn-Teller
distortion \cite{Muya2012111} which was found to yield a slightly more stable $\mathrm{T}_{\mathrm{h}}$ 20-0 fullerene.
This distortion has no impact on the conclusion of this work. The associated puckering of the filling atoms does
not impact the global pyramidalization angle because they are alternatively moved inside and outside the structure.
Alternatively, this distortion can be understood in the IFPR since for the 20-0 structure the filling atoms are
the most reactive sites. In the rest of the text, the fullerenes are referenced according to their symmetry
and their number of filled hexagons and pentagons \cite{PhysRevB.83.081403}. For example, the first proposed fullerene \cite{PhysRevLett.98.166804} becomes
the $\mathrm{I}_{\mathrm{h}}$ 20-0 and the volleyball structure \cite{PhysRevB.82.153409} becomes the
$\mathrm{T}_{\mathrm{h}}$ 8-12.

\section{Results and Discussion}

\subsection{Effect of filled pentagons on pure borofullerenes}
In the fullerene structure, the role of the pentagons is to allow for curvature by increasing
the pyramidalization of its corners \cite{Haddon17091993}. Hence, most of the strain is stored in the pentagons.
In carbon, this leads to the isolated pentagon rule (IPR) \cite{IPR,doi:10.1021/cr990332q} which is a rather strong 
requirement on the stability of pure neutral carbon fullerenes. In the boron fullerenes, the pentagons
essentially play the same role. The only caveat is that the filling boron atoms can be placed to lower the global strain
of the fullerene by increasing their pyramidalization angle. Since these atoms serve mostly as dopants
they are not constrained to a $sp^2$ bonding pattern, in a way reminiscent of the carbon atoms in $\eta^2$-complexes
\cite{Haddon17091993} of carbon fullerenes, and they do not contribute to the strain. The strain is characterized by
the average pyramidalization angle (see Methods) and is presented in Figure \ref{Fig:systems}. As can be seen, the inclusion
of filled pentagons always releases strain and the energetically preferred structure ($\mathrm{D}_{3\mathrm{d}}$ 14-6) sports
the smallest angle of all the isomers. 

Nevertheless, not all the isomers are stabilized when compared to the $\mathrm{I}_{\mathrm{h}}$ 20-0. Indeed, the position of the
20 filling atoms with respect to the 60 atoms backbone cage appears to be crucial.
They should be distributed such that the self-doping rule for boron $sp^2$ systems is satisfied: filled rings should be
bordered by alternating filled and empty neighbor rings. Of course, this cannot be realized for the filled pentagons and
they will thus inevitably be frustrated. The most stable local environment for filled pentagons, presented in Figure
\ref{Fig:pentagon_bonding}, is the closest one can find to the self-doping requirement: almost all backbone atoms contribute half their electrons to the
pentagon, except the atom located at the junction of the two filled neighbors which contributes only one. This atom has a coordination number of six, instead
of a coordination of five which is normally found in the backbone. It thus behaves more like a dopant atom than like a backbone atom and we can see this
side of the pentagon has three inter-penetrating filled-hexagons (see supplementary information). It is this boron atom which sports the largest
pyramidalization angle but because of its almost dopant nature, it does not contribute to the strain.

These two considerations (strain and self-doping) explain the total energy differences observed in Figure \ref{Fig:systems}.
The inclusion of only one or two filled pentagons always results in a destabilized fullerene because the self-doping rule
cannot be achieved. The first stabilized fullerene is the $\mathrm{C}_{3}$ 17-3, where the three-filled pentagons are
clustered around one hexagon which becomes the pole of symmetry. All other stabilized structures, are completely formed
of pentagons with the same local environment, except the $\mathrm{D}_{3\mathrm{d}}$ 10-10 which
also possess other types of pentagons but which are placed to complement each other. This demonstrates that such an environment is the most stable for the
$\mathrm{B}_{80}$ and that the increased stability in these isomers stems mostly by global release of strain as
denoted by their decreased average pyramidalization angles. This balance between strain and electronic issues form a
generalization of the IPR, which we call the isolated filled pentagons (IFPR) rule, in the borofullerenes.


\begin{figure}
\begin{center}
\includegraphics[scale=0.175]{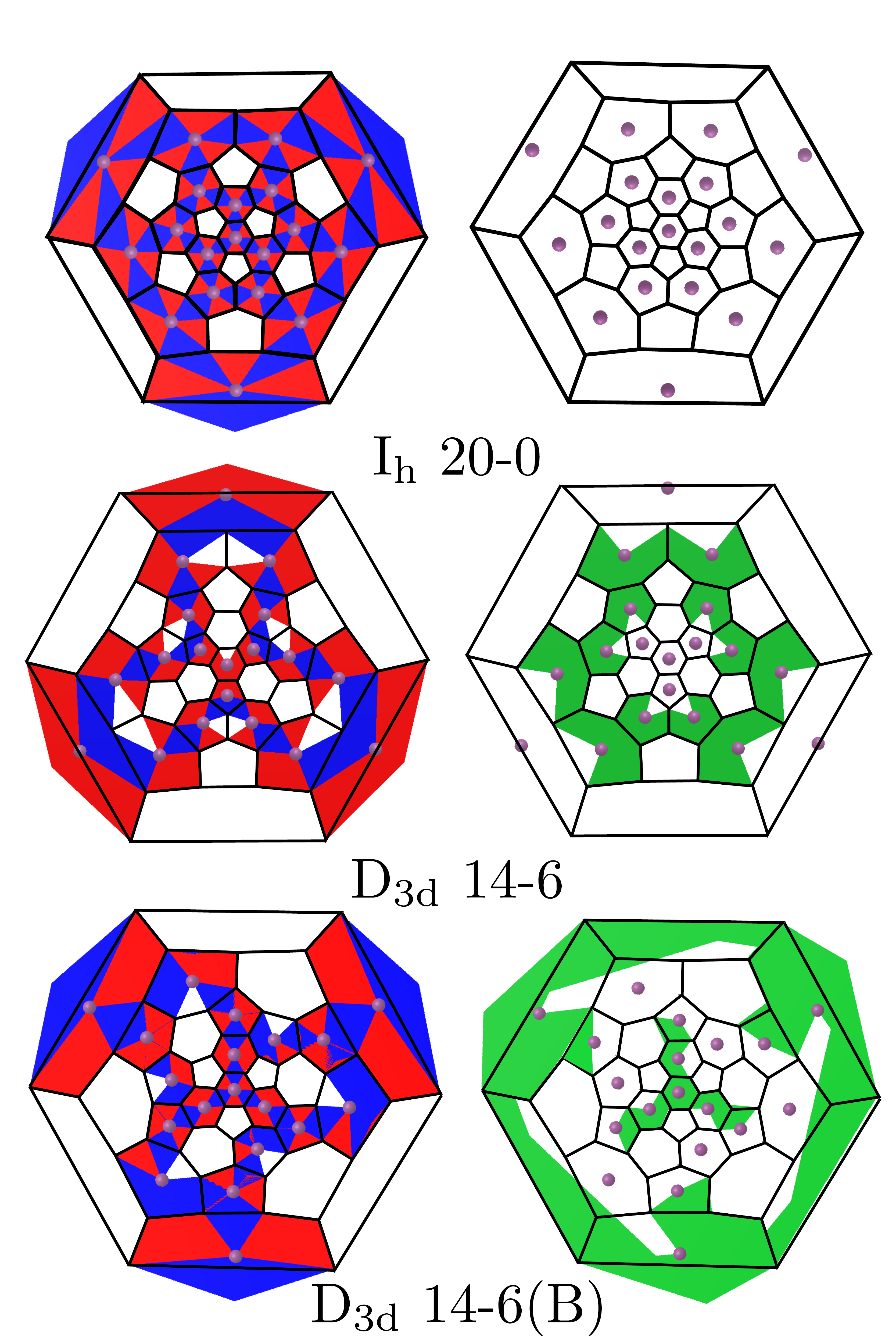}
\caption{Schlegel diagrams with superposed Wannier states for the
$\mathrm{I}_{\mathrm{h}}$ 20-0 and both $\mathrm{C}_{3\mathrm{v}}$ 14-6
structures. The 3c-2e bonds are in red and the 4c-2e bonds are in blue. The
delocalized $\pi_z$ bonds are displayed in green on the right panels. For clarity, the filling
atoms are shown as big purple circles.
The $\mathrm{I}_{\mathrm{h}}$ 20-0 is formed of alternating 3c-2e bonds and 4c-2e bonds
in the same K\'ekul\'e structure then the one observed in $C_{60}$. In the case of the
$\mathrm{D}_{3\mathrm{d}}$ 14-6, we can see a complete percolation of the
$\pi_z$ states on the equator. For the $\mathrm{D}_{3\mathrm{d}}$ 14-6(B), the
$\pi_z$ states span out of the poles in three lightning shaped arms that do not
meet.} 
\label{Fig:stereo}
\end{center}
\end{figure}

\subsection{Partial aromaticity and reactivity}

To quantify the $sp^2$ character of the cages, we perform a decomposition of its orbitals in terms 
of localized bonds via the Wannier transformation \cite{Mostofi2008685} (see Methods section). 
The K\'ekul\'e bonding pattern found in the original $\mathrm{I}_{\mathrm{h}}$ 20-0 is
similar to the one found in $\mathrm{C}_{60}$ except for the multi-center character of the 
bonds: there is a double four-center-two-electron (4c-2e) $\sigma$-bond located between two
adjacent filled-hexagons and a three-center-two-electron (3c-2e) $\sigma$-bond
located on the edge of the empty pentagons, as seen on the Schlegel diagrams of Figure \ref{Fig:stereo}.
This is also similar to the bonding pattern found in the boron
$\alpha$-sheet \cite{C1CP20439D}, with one notable exception: the $\pi_z$ states are now
partly localized inside the double 4c-2e bonds. This lowering of the aromaticity \cite{C1CP22521A} further
destabilizes fullerene structures with respect to their planar counterparts.

For the other isomers, the bonding pattern of the filled pentagons slightly changes because of the differing symmetry,
i.e. not all surrounding hexagons are necessarily equivalent. Nevertheless, a Wannier analysis shows that most states
remain the same. Notable changes stems from local release of strain which are associated 
to the presence of  strongly delocalized $\pi_z$ 7c-2e states. As can be seen in Figure \ref{Fig:stereo}, in the
$\mathrm{D}_{3\mathrm{d}}$ 14-6 these delocalized $\pi_z$ electrons percolate around the equator of the
isomer. This equatorial aromaticity is analogous to the one observed in $C_{70}$, albeit weaker : the equatorial angles are
of $10.5^\circ$ for this isomer while they are of $8.75^\circ$ \cite{Haddon17091993} for $C_{70}$.
Consequently, the atoms of the equator are thus expected to be less reactive than their $\mathrm{I}_{\mathrm{h}}$
20-0 counterparts. At the same time, other sections of this isomer seem far more prone to reactions.
The chemical hardness, determined by the HOMO-LUMO gap, can be used on the DFT level as
a good representation of the reactivity trends in these isomers. Furthermore, the local
strain coupled with the atomic contributions to frontier orbitals was shown to
yield accurate reactivity maps for convex clusters \cite{doi:10.1021/jp304891e}. The local strain on the
backbone atoms of the poles ($11.92^\circ$) is larger than for the
$\mathrm{I}_{\mathrm{h}}$ 20-0 ($11.46^\circ$) which leads to a higher strain activated reactivity.
The non-degenerate LUMO level of this isomer is also located near these atoms suggesting that they indeed
represents the preferred sites for reduction reactions.

One can thus wonder if a different distribution of the same number of filled pentagons can lead to a decreased
reactivity while maintaining the global release of strain. The IFPR suggests that this is not possible because
release of global strain is associated with its local increase in some zone of the structure. Thus increased stability
comes at the price of increased reactivity. This is demonstrated within a second $\mathrm{D}_{3\mathrm{d}}$ 14-6(B)
isomer. This structure is formed with exactly the same pentagonal environment but the filled
pentagons now lie on the equator.  
This corresponds to the maximally separated uniform distribution of filled pentagons. In this case, both the global pyramidalization angle ($11.35^\circ$)
and the minimal angle ($10.60^\circ$) are higher than the ones of the $\mathrm{D}_{3\mathrm{d}}$ 14-6. Now, its HOMO-LUMO gap is almost five times higher
suggesting a much lower reactivity. An inspection of its doubly degenerated LUMO reveals that it is now delocalized on the equator, suggesting that it
possesses less specific reactive sites. The same kind of delocalized $\pi_z$ electrons is seen in this isomer,
but because of the new pentagonal arrangement, they are this time arching from the poles in three lightning shaped arms that
terminate on the filled pentagons (represented in the lower right section of Figure \ref{Fig:stereo}). This points to a
lower reactivity of the poles. These $\pi_z$ states do not percolate leading to a lower resonance energy then
the $\mathrm{D}_{3\mathrm{d}}$ 14-6.  The increased strain and reduced resonance energy explain the energy
difference between the two 14-6 isomers.

 \begin{table}
 \begin{tabular}{|c|c|c|c|c|} \cline{2-5}
  \multicolumn{1}{c|}{}              & \multicolumn{2}{|c|}{$\mathrm{Sc}_3$N@$\mathrm{B}_{80}$} &
\multicolumn{2}{|c|}{$\mathrm{B}^{-2}_{80}$}      \\ \cline{2-5}
  \multicolumn{1}{c|}{}              & $\Delta$E [eV]  & $E_{G}$ [eV]                           & $\Delta$E [eV]  & $E_{G}$ [eV] \\ \hline
  $\mathrm{I}_{\mathrm{h}}$ 20-0     &  -10.859        & 0.592                                  &  -3.563         & 0.132        \\ \hline
  $\mathrm{D}_{3\mathrm{d}}$ 14-6    &  -10.973        & 0.229                                  &  -4.591         & 0.550        \\ \hline
  $\mathrm{D}_{3\mathrm{d}}$ 14-6(B) &  -12.976        & 0.385                                  &  -4.286         & 0.175        \\ \hline
 \end{tabular} 
 \caption{Binding energies and HOMO-LUMO separation for the endohedral
 $\mathrm{Sc}_3$N@$\mathrm{B}_{80}$ fullerene and the $\mathrm{B}^{-2}_{80}$ ion.
 Binding energies are mesured with respect to the energy of the pristine cage
 (plus the energy of the added cluster).}
 \label{Table:charged}
 \end{table}

\subsection{Exterior v.s. self-doping}

Having identified the consequences of strain and self-doping, we can now address the influence
of exterior doping, i.e. the interaction with non boron atoms. As mentioned earlier, metallic
 clusters  are envisioned to act as growth centers for endohedral metalloborofullerenes. 
The induced stabilization of each endohedral isomer will depend greatly on the nature, symmetry and orientation of
the seed. Thus, the identification of a good candidate for synthesis would necessitate a complete zoology of metallic
clusters coupled with their effect on different isomers. This is beyond the scope of this paper. We will rather focus on
the general effects of the inclusion of the metallic clusters on the pentagonal unit, 
by considering two basic examples: the $\mathrm{Sc}_3$N cluster and the $\mathrm{B}^{-2}_{80}$ anions.

As can be seen in Table \ref{Table:charged}, all isomers are stabilized by the inclusion of the $\mathrm{Sc}_3$N ``seed''.

For the $\mathrm{I}_{\mathrm{h}}$ 20-0, we find that the nitrogen atom departs from the plane formed by the Sc to form a covalent bond with a filling boron atom, thus
reproducing the results of Peng et al. \cite{doi:10.1021/jp9019848}.  This is consistent with the reactivity picture
described by the LUMO and the strain. A Bader analysis of the electron density reveals that the Sc atoms have lost
1.5 electrons while the N atom gained 2.5 electrons. This corresponds to a net transfer of 2 electrons to the boron
system. A Jahn-Teller deformation is observed at this filling since the LUMO of the $\mathrm{I}_{\mathrm{h}}$ 20-0 is threefold
degenerate. This reduces the symmetry of the system to $C_1$ leading to a decrease of the HOMO-LUMO gap of the system to 0.592 eV. The global pyramidalization angle
remained the same ($11.46^\circ$) showing that the endohedral cluster in the $\mathrm{I}_{\mathrm{h}}$ 20-0 do not impact the global strain.
Since the symmetry is decreased, there is now a greater scatter in the local strain leading to some preferred reactive sites which are located close to the N-B bond.

In the $\mathrm{D}_{3\mathrm{d}}$ 14-6 isomers, things are quite different. The metallic cluster adopts
a quasi-planar structure which places itself on the plane containing the equator. This shows that no covalent
bonds between the cluster and the fullerenes are formed, which is confirmed by the electronic density. This also agrees with the
previous reactivity picture: the LUMO of these systems is formed of backbone $\pi_z$ states. Interior covalent bonding
with these atoms is therefore unlikely: it inevitably leads to an increase of the global strain which greatly destabilizes
these states. This is the origin of the ``topological'' protection invoked in the literature \cite{PhysRevB.83.081403}.
Nevertheless, the global pyramidalization angle of the $\mathrm{D}_{3\mathrm{d}}$ 14-6(B) was greatly reduced ($11.09^\circ$)
and is now similar with the $\mathrm{D}_{3\mathrm{d}}$ 14-6.
Consequently, as we can see in Table \ref{Table:charged}, the stability of the $\mathrm{D}_{3\mathrm{d}}$ 14-6(B)
is greatly increased. It becomes the most energetically stable isomer. This can be traced to electrostatic effects. A
Bader analysis of the charge density displays a 3.2 electron transfer from the cluster to the cages
(with +1.6e Sc atoms and -1.6e N atom). These extra electrons are mostly clustered around the equator which maximizes
the electrostatic interaction with the positively charged Sc atoms. Its increased stability
is thus a consequence of the planar geometry of the seed. 
By maximizing the electrostatic interactions, by varying the shape and the symmetry of the metallic clusters,
it is thus possible to maximize the driving force which will select a given fullerene structure during growth.

Interestingly, the seed increases the HOMO-LUMO gap of the $\mathrm{D}_{3\mathrm{d}}$ 14-6 suggesting a decrease in reactivity. This can be maximized 
if the metallic cluster is chosen to yield an electron count that matches the distribution of the electronic levels of the neutral fullerene.
To this end, we have studied the corresponding $\mathrm{B}^{-2}_{80}$ ions. Here, we see that the stabilization
caused by the population of the $\pi_{z}$ electrons is maximal for the $\mathrm{D}_{3\mathrm{d}}$ 14-6 which stays the most energetically
favored isomer. Since the LUMO of this isomer is not degenerate, there is no Jahn-Teller deformation in this system and the pyramidalization angles stays the same.
The rigid band approximation works well in these systems, since a difference of the anion and neutral densities show that the extra charge
is located on the LUMOs and the energy levels are basically unperturbed. The extra electrons are thus distributed in the $\pi$ states of the backbone atoms of the poles of the
$\mathrm{D}_{3\mathrm{d}}$ 14-6 and it is the increased resonance energy which is responsible for the stabilization and decreased reactivity of the structure.
Unsurprisingly, in the case of the $\mathrm{D}_{3\mathrm{d}}$ 14-6(B), there is a Jahn-Teller deformation since in its neutral state the LUMO is twice degenerated.
This deformation allows it to decrease its global pyramidalization angle ($11.27^\circ$) which is responsible for its further stabilization. 

\section{Summary and Conclusions}

To rationalize the effect of exterior and self-doping, we carried out simulations on the $B_{80}$ isomers. This has lead us to formulate
the generalized isolated filled pentagon rule (IFPR) which states that a peculiar environment is necessary for the
stability of the filled pentagons. Because of polymorphism, the IFPR is not a strong constraint on stability,
as the IPR, but rather a guideline for constructing the most stable known structures. It is also shown that
increased $\pi$ conjugation enabled by local decrease of strain plays a large role in the stabilization of
certain isomers. The IFPR isomers are thus reminiscent of $C_{70}$ where equatorial aromaticity plays a stronger role.
This is in good agreement with the simulated ring currents for the $\mathrm{I}_{\mathrm{h}}$ 20-0
\cite{C1CP22521A}. Further study of the effect of filled pentagons on the ring currents of $\mathrm{B}_{80}$ would be needed.
Furthermore, the IFPR clearly indicates that the synthesis of pure borofullerenes faces difficult challenges because
increased stability comes at the price of increased strain induced reactivity. 

The IFPR should be valid for larger boron fullerenes. In this case, the increase of the hexagon to pentagon
ratio enables greater freedom to determine the local environment of the filled pentagons.  It then becomes possible to
maintain a same filling pattern than the $\alpha$-sheet, or other stable sheets, for the hexagons located far from
the pentagons all the while modifying the ones close to the pentagons to maximize stability. However, the relative energy
differences should drop as the size of the fullerene increases since the effect of the curvature diminishes. Larger
fullerenes are thus expected to have a denser energy spectrum recovering the stronger
polymorphism \cite{doi:10.1021/nl3004754} of the boron sheets. 

Self-doping alone is not enough to guarantee the existence of the fullerene structures. It is thus
necessary to explore other means of stabilizing these
structures. The use of endohedral ``seeds'' for cluster growth could provide further doping leading
to greater stabilization.  In an effort to understand if a future synthesis of endohedral borofullerenes
is feasible, we have studied the effect of this exterior doping on the IFPR. Of course, we have restricted
ourself to IPR fullerenes and one must keep in mind that in endohedral carbon fullerenes departure from the
IPR is well-known \cite{non-IPR} and that this can also be the case in endohedral borofullerenes. In the two
cases considered, namely the $\mathrm{Sc}_3$N cluster and the $B^{-2}_{80}$ anions, the stability of the isomers
were systematically increased. The effect of the IFPR was found to inhibit direct reaction with the endohedral
cluster since they would inevitably increase the strain of the fullerene shell. Furthermore, it is shown, that
during growth, the symmetry of the metallic cluster will select a given isomer which will minimize the electrostatic effects.
In other words, the ``seed'' can restore a greater energetically motivated (~2 eV)  driving force.
We have also demonstrated that electron addition can decrease the reactivity of the borofullerenes thus reducing the probability of forming
outwards $sp^3$ structures during synthesis. We thus show that a metallic seed has all the requirements needed to solve the 
difficulties restricting the synthesis of boron fullerenes.



\begin{thebibliography}{99}
\makeatletter
\providecommand \@ifxundefined [1]{%
 \ifx #1\undefined \expandafter \@firstoftwo
 \else \expandafter \@secondoftwo
\fi
}%
\providecommand \@ifnum [1]{%
 \ifnum #1\expandafter \@firstoftwo
 \else \expandafter \@secondoftwo
\fi
}%
\providecommand \enquote [1]{``#1''}%
\providecommand \bibnamefont  [1]{#1}%
\providecommand \bibfnamefont [1]{#1}%
\providecommand \citenamefont [1]{#1}%
\providecommand\href[0]{\@sanitize\@href}%
\providecommand\@href[1]{\endgroup\@@startlink{#1}\endgroup\@@href}%
\providecommand\@@href[1]{#1\@@endlink}%
\providecommand \@sanitize [0]{\begingroup\catcode`\&12\catcode`\#12\relax}%
\@ifxundefined \pdfoutput {\@firstoftwo}{%
 \@ifnum{\z@=\pdfoutput}{\@firstoftwo}{\@secondoftwo}%
}{%
 \providecommand\@@startlink[1]{\leavevmode}%
 \providecommand\@@endlink[0]{}%
}{%
 \providecommand\@@startlink[1]{%
  \leavevmode
  \pdfstartlink
   attr{/Border[0 0 1 ]/H/I/C[0 1 1]}%
   user{/Subtype/Link/A<</Type/Action/S/URI/URI(#1)>>}%
  \relax
 }%
 \providecommand\@@endlink[0]{\pdfendlink}%
}%
\providecommand \url  [0]{\begingroup\@sanitize \@url }%
\providecommand \@url [1]{\endgroup\@href {#1}{\urlprefix}}%
\providecommand \urlprefix [0]{URL }%
\providecommand \Eprint[0]{\href }%
\@ifxundefined \urlstyle {%
  \providecommand \doi [1]{doi:\discretionary{}{}{}#1}%
}{%
  \providecommand \doi [0]{doi:\discretionary{}{}{}\begingroup
  \urlstyle{rm}\Url }%
}%
\providecommand \doibase [0]{http://dx.doi.org/}%
\providecommand \Doi[1]{\href{\doibase#1}}%
\providecommand \bibAnnote [3]{%
  \BibitemShut{#1}%
  \begin{quotation}\noindent
    \textsc{Key:}\ #2\\\textsc{Annotation:}\ #3%
  \end{quotation}%
}%
\providecommand \bibAnnoteFile [2]{%
  \IfFileExists{#2}{\bibAnnote {#1} {#2} {\input{#2}}}{}%
}%
\providecommand \typeout [0]{\immediate \write \m@ne }%
\providecommand \selectlanguage [0]{\@gobble}%
\providecommand \bibinfo [0]{\@secondoftwo}%
\providecommand \bibfield [0]{\@secondoftwo}%
\providecommand \translation [1]{[#1]}%
\providecommand \BibitemOpen[0]{}%
\providecommand \bibitemStop [0]{}%
\providecommand \bibitemNoStop [0]{.\EOS\space}%
\providecommand \EOS [0]{\spacefactor3000\relax}%
\providecommand \BibitemShut [1]{\csname bibitem#1\endcsname}%
\bibitem{doi:10.1021/ar0300266}%
  \BibitemOpen
  \bibfield{author}{%
  \bibinfo {author} {\bibfnamefont{E.~D.}\ \bibnamefont{Jemmis}}\ and\ \bibinfo
  {author} {\bibfnamefont{E.~G.}\ \bibnamefont{Jayasree}},\ }%
  \bibfield{journal}{%
  \bibinfo {journal} {Accounts of Chemical Research}\ }%
  \textbf{\bibinfo {volume} {36}},\ \bibinfo {pages} {816} (\bibinfo {year}
  {2003})%
  \bibAnnoteFile{NoStop}{doi:10.1021/ar0300266}%
\bibitem{doi:10.1021/ja01634a089}%
  \BibitemOpen
  \bibfield{author}{%
  \bibinfo {author} {\bibfnamefont{M.~E.}\ \bibnamefont{Jones}}\ and\ \bibinfo
  {author} {\bibfnamefont{R.~E.}\ \bibnamefont{Marsh}},\ }%
  \bibfield{journal}{%
  \bibinfo {journal} {Journal of the American Chemical Society}\ }%
  \textbf{\bibinfo {volume} {76}},\ \bibinfo {pages} {1434} (\bibinfo {year}
  {1954})%
  \bibAnnoteFile{NoStop}{doi:10.1021/ja01634a089}%
\bibitem{MgB2_nature}%
  \BibitemOpen
  \bibfield{author}{%
  \bibinfo {author} {\bibfnamefont{J.}~\bibnamefont{Nagamatsu}}, \bibinfo
  {author} {\bibfnamefont{N.}~\bibnamefont{Nakagawa}}, \bibinfo {author}
  {\bibfnamefont{T.}~\bibnamefont{Muranaka}}, \bibinfo {author}
  {\bibfnamefont{Y.}~\bibnamefont{Zenitani}},\ and\ \bibinfo {author}
  {\bibfnamefont{J.}~\bibnamefont{Akimitsu}},\ }%
  \bibfield{journal}{%
  \bibinfo {journal} {Nature}\ }%
  \textbf{\bibinfo {volume} {410}},\ \bibinfo {pages} {63} (\bibinfo {year}
  {2001})%
  \bibAnnoteFile{NoStop}{MgB2_nature}%
\bibitem{PhysRevB.80.134113}%
  \BibitemOpen
  \bibfield{author}{%
  \bibinfo {author} {\bibfnamefont{H.}~\bibnamefont{Tang}}\ and\ \bibinfo
  {author} {\bibfnamefont{S.}~\bibnamefont{Ismail-Beigi}},\ }%
  \bibfield{journal}{%
  \bibinfo {journal} {Phys. Rev. B}\ }%
  \textbf{\bibinfo {volume} {80}},\ \bibinfo {pages} {134113} (\bibinfo {year}
  {2009})%
  \bibAnnoteFile{NoStop}{PhysRevB.80.134113}%
\bibitem{nanosheet}%
  \BibitemOpen
  \bibfield{author}{%
  \bibinfo {author} {\bibfnamefont{H.-J.}\ \bibnamefont{Zhai}}, \bibinfo
  {author} {\bibfnamefont{B.}~\bibnamefont{Kiran}},\ and\ \bibinfo {author}
  {\bibfnamefont{W.}~\bibnamefont{Lai-Sheng}},\ }%
  \bibfield{journal}{%
  \bibinfo {journal} {Nature Materials}\ }%
  \textbf{\bibinfo {volume} {2}},\ \bibinfo {pages} {827} (\bibinfo {year}
  {2003})%
  \bibAnnoteFile{NoStop}{nanosheet}%
\bibitem{nanosheet2}%
  \BibitemOpen
  \bibfield{author}{%
  \bibinfo {author} {\bibfnamefont{W.}~\bibnamefont{Huang}}, \bibinfo {author}
  {\bibfnamefont{A.~P.}\ \bibnamefont{Sergeeva}}, \bibinfo {author}
  {\bibfnamefont{H.-J.}\ \bibnamefont{Zhai}}, \bibinfo {author}
  {\bibfnamefont{B.~B.}\ \bibnamefont{Averkiev}}, \bibinfo {author}
  {\bibfnamefont{L.-S.}\ \bibnamefont{Wang}},\ and\ \bibinfo {author}
  {\bibfnamefont{A.~I.}\ \bibnamefont{Boldyrev}},\ }%
  \bibfield{journal}{%
  \bibinfo {journal} {Nature Chemistry}\ }%
  \textbf{\bibinfo {volume} {2}},\ \bibinfo {pages} {202} (\bibinfo {year}
  {2010})%
  \bibAnnoteFile{NoStop}{nanosheet2}%
\bibitem{Boustani1997182}%
  \BibitemOpen
  \bibfield{author}{%
  \bibinfo {author} {\bibfnamefont{I.}~\bibnamefont{Boustani}},\ }%
  \bibfield{journal}{%
  \bibinfo {journal} {Journal of Solid State Chemistry}\ }%
  \textbf{\bibinfo {volume} {133}},\ \bibinfo {pages} {182 } (\bibinfo {year}
  {1997})%
  \bibAnnoteFile{NoStop}{Boustani1997182}%
\bibitem{PhysRevLett.99.115501}%
  \BibitemOpen
  \bibfield{author}{%
  \bibinfo {author} {\bibfnamefont{H.}~\bibnamefont{Tang}}\ and\ \bibinfo
  {author} {\bibfnamefont{S.}~\bibnamefont{Ismail-Beigi}},\ }%
  \bibfield{journal}{%
  \bibinfo {journal} {Phys. Rev. Lett.}\ }%
  \textbf{\bibinfo {volume} {99}},\ \bibinfo {pages} {115501} (\bibinfo {year}
  {2007})%
  \bibAnnoteFile{NoStop}{PhysRevLett.99.115501}%
\bibitem{PhysRevB.74.035413}%
  \BibitemOpen
  \bibfield{author}{%
  \bibinfo {author} {\bibfnamefont{J.}~\bibnamefont{Kunstmann}}\ and\ \bibinfo
  {author} {\bibfnamefont{A.}~\bibnamefont{Quandt}},\ }%
  \bibfield{journal}{%
  \bibinfo {journal} {Phys. Rev. B}\ }%
  \textbf{\bibinfo {volume} {74}},\ \bibinfo {pages} {035413} (\bibinfo {year}
  {2006})%
  \bibAnnoteFile{NoStop}{PhysRevB.74.035413}%
\bibitem{doi:10.1021/nl073295o}%
  \BibitemOpen
  \bibfield{author}{%
  \bibinfo {author} {\bibfnamefont{A.~K.}\ \bibnamefont{Singh}}, \bibinfo
  {author} {\bibfnamefont{A.}~\bibnamefont{Sadrzadeh}},\ and\ \bibinfo {author}
  {\bibfnamefont{B.~I.}\ \bibnamefont{Yakobson}},\ }%
  \bibfield{journal}{%
  \bibinfo {journal} {Nano Letters}\ }%
  \textbf{\bibinfo {volume} {8}},\ \bibinfo {pages} {1314} (\bibinfo {year}
  {2008})%
  \bibAnnoteFile{NoStop}{doi:10.1021/nl073295o}%
\bibitem{Zope2011193}%
  \BibitemOpen
  \bibfield{author}{%
  \bibinfo {author} {\bibfnamefont{R.~R.}\ \bibnamefont{Zope}}\ and\ \bibinfo
  {author} {\bibfnamefont{T.}~\bibnamefont{Baruah}},\ }%
  \bibfield{journal}{%
  \bibinfo {journal} {Chemical Physics Letters}\ }%
  \textbf{\bibinfo {volume} {501}},\ \bibinfo {pages} {193 } (\bibinfo {year}
  {2011})%
  \bibAnnoteFile{NoStop}{Zope2011193}%
\bibitem{doi:10.1021/nl3004754}%
  \BibitemOpen
  \bibfield{author}{%
  \bibinfo {author} {\bibfnamefont{E.~S.}\ \bibnamefont{Penev}}, \bibinfo
  {author} {\bibfnamefont{S.}~\bibnamefont{Bhowmick}}, \bibinfo {author}
  {\bibfnamefont{A.}~\bibnamefont{Sadrzadeh}},\ and\ \bibinfo {author}
  {\bibfnamefont{B.~I.}\ \bibnamefont{Yakobson}},\ }%
  \bibfield{journal}{%
  \bibinfo {journal} {Nano Letters}\ }%
  \textbf{\bibinfo {volume} {12}},\ \bibinfo {pages} {2441} (\bibinfo {year}
  {2012})%
  \bibAnnoteFile{NoStop}{doi:10.1021/nl3004754}%
\bibitem{doi:10.1021/nn302696v}%
  \BibitemOpen
  \bibfield{author}{%
  \bibinfo {author} {\bibfnamefont{X.}~\bibnamefont{Wu}}, \bibinfo {author}
  {\bibfnamefont{J.}~\bibnamefont{Dai}}, \bibinfo {author}
  {\bibfnamefont{Y.}~\bibnamefont{Zhao}}, \bibinfo {author}
  {\bibfnamefont{Z.}~\bibnamefont{Zhuo}}, \bibinfo {author}
  {\bibfnamefont{J.}~\bibnamefont{Yang}},\ and\ \bibinfo {author}
  {\bibfnamefont{X.~C.}\ \bibnamefont{Zeng}},\ }%
  \bibfield{journal}{%
  \bibinfo {journal} {ACS Nano}\ }%
  \textbf{\bibinfo {volume} {6}},\ \bibinfo {pages} {7443} (\bibinfo {year}
  {2012})%
  \bibAnnoteFile{NoStop}{doi:10.1021/nn302696v}%
\bibitem{PhysRevLett.106.225502}%
  \BibitemOpen
  \bibfield{author}{%
  \bibinfo {author} {\bibfnamefont{S.}~\bibnamefont{De}}, \bibinfo {author}
  {\bibfnamefont{A.}~\bibnamefont{Willand}}, \bibinfo {author}
  {\bibfnamefont{M.}~\bibnamefont{Amsler}}, \bibinfo {author}
  {\bibfnamefont{P.}~\bibnamefont{Pochet}}, \bibinfo {author}
  {\bibfnamefont{L.}~\bibnamefont{Genovese}},\ and\ \bibinfo {author}
  {\bibfnamefont{S.}~\bibnamefont{Goedecker}},\ }%
  \bibfield{journal}{%
  \bibinfo {journal} {Phys. Rev. Lett.}\ }%
  \textbf{\bibinfo {volume} {106}},\ \bibinfo {pages} {225502} (\bibinfo {year}
  {2011})%
  \bibAnnoteFile{NoStop}{PhysRevLett.106.225502}%
\bibitem{C1CP20439D}%
  \BibitemOpen
  \bibfield{author}{%
  \bibinfo {author} {\bibfnamefont{T.~R.}\ \bibnamefont{Galeev}}, \bibinfo
  {author} {\bibfnamefont{Q.}~\bibnamefont{Chen}}, \bibinfo {author}
  {\bibfnamefont{J.-C.}\ \bibnamefont{Guo}}, \bibinfo {author}
  {\bibfnamefont{H.}~\bibnamefont{Bai}}, \bibinfo {author}
  {\bibfnamefont{C.-Q.}\ \bibnamefont{Miao}}, \bibinfo {author}
  {\bibfnamefont{H.-G.}\ \bibnamefont{Lu}}, \bibinfo {author}
  {\bibfnamefont{A.~P.}\ \bibnamefont{Sergeeva}}, \bibinfo {author}
  {\bibfnamefont{S.-D.}\ \bibnamefont{Li}},\ and\ \bibinfo {author}
  {\bibfnamefont{A.~I.}\ \bibnamefont{Boldyrev}},\ }%
  \bibfield{journal}{%
  \bibinfo {journal} {Phys. Chem. Chem. Phys.}\ }%
  \textbf{\bibinfo {volume} {13}},\ \bibinfo {pages} {11575} (\bibinfo {year}
  {2011})%
  \bibAnnoteFile{NoStop}{C1CP20439D}%
\bibitem{PhysRevLett.98.166804}%
  \BibitemOpen
  \bibfield{author}{%
  \bibinfo {author} {\bibfnamefont{N.}~\bibnamefont{Gonzalez~Szwacki}},
  \bibinfo {author} {\bibfnamefont{A.}~\bibnamefont{Sadrzadeh}},\ and\ \bibinfo
  {author} {\bibfnamefont{B.~I.}\ \bibnamefont{Yakobson}},\ }%
  \bibfield{journal}{%
  \bibinfo {journal} {Phys. Rev. Lett.}\ }%
  \textbf{\bibinfo {volume} {98}},\ \bibinfo {pages} {166804} (\bibinfo {year}
  {2007})%
  \bibAnnoteFile{NoStop}{PhysRevLett.98.166804}%
\bibitem{PhysRevB.79.161403}%
  \BibitemOpen
  \bibfield{author}{%
  \bibinfo {author} {\bibfnamefont{R.~R.}\ \bibnamefont{Zope}}, \bibinfo
  {author} {\bibfnamefont{T.}~\bibnamefont{Baruah}}, \bibinfo {author}
  {\bibfnamefont{K.~C.}\ \bibnamefont{Lau}}, \bibinfo {author}
  {\bibfnamefont{A.~Y.}\ \bibnamefont{Liu}}, \bibinfo {author}
  {\bibfnamefont{M.~R.}\ \bibnamefont{Pederson}},\ and\ \bibinfo {author}
  {\bibfnamefont{B.~I.}\ \bibnamefont{Dunlap}},\ }%
  \bibfield{journal}{%
  \bibinfo {journal} {Phys. Rev. B}\ }%
  \textbf{\bibinfo {volume} {79}},\ \bibinfo {pages} {161403} (\bibinfo {year}
  {2009})%
  \bibAnnoteFile{NoStop}{PhysRevB.79.161403}%
\bibitem{PhysRevB.78.201401}%
  \BibitemOpen
  \bibfield{author}{%
  \bibinfo {author} {\bibfnamefont{Q.-B.}\ \bibnamefont{Yan}}, \bibinfo
  {author} {\bibfnamefont{X.-L.}\ \bibnamefont{Sheng}}, \bibinfo {author}
  {\bibfnamefont{Q.-R.}\ \bibnamefont{Zheng}}, \bibinfo {author}
  {\bibfnamefont{L.-Z.}\ \bibnamefont{Zhang}},\ and\ \bibinfo {author}
  {\bibfnamefont{G.}~\bibnamefont{Su}},\ }%
  \bibfield{journal}{%
  \bibinfo {journal} {Phys. Rev. B}\ }%
  \textbf{\bibinfo {volume} {78}},\ \bibinfo {pages} {201401} (\bibinfo {year}
  {2008})%
  \bibAnnoteFile{NoStop}{PhysRevB.78.201401}%
\bibitem{PhysRevB.82.153409}%
  \BibitemOpen
  \bibfield{author}{%
  \bibinfo {author} {\bibfnamefont{X.-Q.}\ \bibnamefont{Wang}},\ }%
  \bibfield{journal}{%
  \bibinfo {journal} {Phys. Rev. B}\ }%
  \textbf{\bibinfo {volume} {82}},\ \bibinfo {pages} {153409} (\bibinfo {year}
  {2010})%
  \bibAnnoteFile{NoStop}{PhysRevB.82.153409}%
\bibitem{PhysRevB.83.081403}%
  \BibitemOpen
  \bibfield{author}{%
  \bibinfo {author} {\bibfnamefont{P.}~\bibnamefont{Pochet}}, \bibinfo {author}
  {\bibfnamefont{L.}~\bibnamefont{Genovese}}, \bibinfo {author}
  {\bibfnamefont{S.}~\bibnamefont{De}}, \bibinfo {author}
  {\bibfnamefont{S.}~\bibnamefont{Goedecker}}, \bibinfo {author}
  {\bibfnamefont{D.}~\bibnamefont{Caliste}}, \bibinfo {author}
  {\bibfnamefont{S.~A.}\ \bibnamefont{Ghasemi}}, \bibinfo {author}
  {\bibfnamefont{K.}~\bibnamefont{Bao}},\ and\ \bibinfo {author}
  {\bibfnamefont{T.}~\bibnamefont{Deutsch}},\ }%
  \bibfield{journal}{%
  \bibinfo {journal} {Phys. Rev. B}\ }%
  \textbf{\bibinfo {volume} {83}},\ \bibinfo {pages} {081403(R)} (\bibinfo {year}
  {2011})%
  \bibAnnoteFile{NoStop}{PhysRevB.83.081403}%
\bibitem{li:074302}%
  \BibitemOpen
  \bibfield{author}{%
  \bibinfo {author} {\bibfnamefont{F.}~\bibnamefont{Li}}, \bibinfo {author}
  {\bibfnamefont{P.}~\bibnamefont{Jin}}, \bibinfo {author}
  {\bibfnamefont{D.}~\bibnamefont{en~Jiang}}, \bibinfo {author}
  {\bibfnamefont{L.}~\bibnamefont{Wang}}, \bibinfo {author}
  {\bibfnamefont{S.~B.}\ \bibnamefont{Zhang}}, \bibinfo {author}
  {\bibfnamefont{J.}~\bibnamefont{Zhao}},\ and\ \bibinfo {author}
  {\bibfnamefont{Z.}~\bibnamefont{Chen}},\ }%
  \bibfield{journal}{%
  \bibinfo {journal} {The Journal of Chemical Physics}\ }%
  \textbf{\bibinfo {volume} {136}},\ \bibinfo {eid} {074302} (\bibinfo {year}
  {2012})%
  \bibAnnoteFile{NoStop}{li:074302}%
\bibitem{PhysRevLett.100.165504}%
  \BibitemOpen
  \bibfield{author}{%
  \bibinfo {author} {\bibfnamefont{D.~L. V.~K.}\ \bibnamefont{Prasad}}\ and\
  \bibinfo {author} {\bibfnamefont{E.~D.}\ \bibnamefont{Jemmis}},\ }%
  \bibfield{journal}{%
  \bibinfo {journal} {Phys. Rev. Lett.}\ }%
  \textbf{\bibinfo {volume} {100}},\ \bibinfo {pages} {165504} (\bibinfo {year}
  {2008})%
  \bibAnnoteFile{NoStop}{PhysRevLett.100.165504}%
\bibitem{C002954H}%
  \BibitemOpen
  \bibfield{author}{%
  \bibinfo {author} {\bibfnamefont{H.}~\bibnamefont{Li}}, \bibinfo {author}
  {\bibfnamefont{N.}~\bibnamefont{Shao}}, \bibinfo {author}
  {\bibfnamefont{B.}~\bibnamefont{Shang}}, \bibinfo {author}
  {\bibfnamefont{L.-F.}\ \bibnamefont{Yuan}}, \bibinfo {author}
  {\bibfnamefont{J.}~\bibnamefont{Yang}},\ and\ \bibinfo {author}
  {\bibfnamefont{X.~C.}\ \bibnamefont{Zeng}},\ }%
  \bibfield{journal}{%
  \bibinfo {journal} {Chem. Commun.}\ }%
  \textbf{\bibinfo {volume} {46}},\ \bibinfo {pages} {3878} (\bibinfo {year}
  {2010})%
  \bibAnnoteFile{NoStop}{C002954H}%
\bibitem{doi:10.1021/jp1018873}%
  \BibitemOpen
  \bibfield{author}{%
  \bibinfo {author} {\bibfnamefont{J.}~\bibnamefont{Zhao}}, \bibinfo {author}
  {\bibfnamefont{L.}~\bibnamefont{Wang}}, \bibinfo {author}
  {\bibfnamefont{F.}~\bibnamefont{Li}},\ and\ \bibinfo {author}
  {\bibfnamefont{Z.}~\bibnamefont{Chen}},\ }%
  \bibfield{journal}{%
  \bibinfo {journal} {The Journal of Physical Chemistry A}\ }%
  \textbf{\bibinfo {volume} {114}},\ \bibinfo {pages} {9969} (\bibinfo {year}
  {2010})%
  \bibAnnoteFile{NoStop}{doi:10.1021/jp1018873}%
\bibitem{wang:133102}%
  \BibitemOpen
  \bibfield{author}{%
  \bibinfo {author} {\bibfnamefont{J.-T.}\ \bibnamefont{Wang}}, \bibinfo
  {author} {\bibfnamefont{C.}~\bibnamefont{Chen}}, \bibinfo {author}
  {\bibfnamefont{E.~G.}\ \bibnamefont{Wang}}, \bibinfo {author}
  {\bibfnamefont{D.-S.}\ \bibnamefont{Wang}}, \bibinfo {author}
  {\bibfnamefont{H.}~\bibnamefont{Mizuseki}},\ and\ \bibinfo {author}
  {\bibfnamefont{Y.}~\bibnamefont{Kawazoe}},\ }%
  \bibfield{journal}{%
  \bibinfo {journal} {Applied Physics Letters}\ }%
  \textbf{\bibinfo {volume} {94}},\ \bibinfo {eid} {133102} (\bibinfo {year}
  {2009})%
  \bibAnnoteFile{NoStop}{wang:133102}%
\bibitem{doi:10.1021/jp9019848}%
  \BibitemOpen
  \bibfield{author}{%
  \bibinfo {author} {\bibfnamefont{P.}~\bibnamefont{Jin}}, \bibinfo {author}
  {\bibfnamefont{C.}~\bibnamefont{Hao}}, \bibinfo {author}
  {\bibfnamefont{Z.}~\bibnamefont{Gao}}, \bibinfo {author}
  {\bibfnamefont{S.~B.}\ \bibnamefont{Zhang}},\ and\ \bibinfo {author}
  {\bibfnamefont{Z.}~\bibnamefont{Chen}},\ }%
  \bibfield{journal}{%
  \bibinfo {journal} {The Journal of Physical Chemistry A}\ }%
  \textbf{\bibinfo {volume} {113}},\ \bibinfo {pages} {11613} (\bibinfo {year}
  {2009})%
  \bibAnnoteFile{NoStop}{doi:10.1021/jp9019848}%
\bibitem{doi:10.1021/jp049301b}%
  \BibitemOpen
  \bibfield{author}{%
  \bibinfo {author} {\bibfnamefont{D.}~\bibnamefont{Ciuparu}}, \bibinfo
  {author} {\bibfnamefont{R.~F.}\ \bibnamefont{Klie}}, \bibinfo {author}
  {\bibfnamefont{Y.}~\bibnamefont{Zhu}},\ and\ \bibinfo {author}
  {\bibfnamefont{L.}~\bibnamefont{Pfefferle}},\ }%
  \bibfield{journal}{%
  \bibinfo {journal} {The Journal of Physical Chemistry B}\ }%
  \textbf{\bibinfo {volume} {108}},\ \bibinfo {pages} {3967} (\bibinfo {year}
  {2004})%
  \bibAnnoteFile{NoStop}{doi:10.1021/jp049301b}%
\bibitem{B919260C}%
  \BibitemOpen
  \bibfield{author}{%
  \bibinfo {author} {\bibfnamefont{F.}~\bibnamefont{Liu}}, \bibinfo {author}
  {\bibfnamefont{C.}~\bibnamefont{Shen}}, \bibinfo {author}
  {\bibfnamefont{Z.}~\bibnamefont{Su}}, \bibinfo {author}
  {\bibfnamefont{X.}~\bibnamefont{Ding}}, \bibinfo {author}
  {\bibfnamefont{S.}~\bibnamefont{Deng}}, \bibinfo {author}
  {\bibfnamefont{J.}~\bibnamefont{Chen}}, \bibinfo {author}
  {\bibfnamefont{N.}~\bibnamefont{Xu}},\ and\ \bibinfo {author}
  {\bibfnamefont{H.}~\bibnamefont{Gao}},\ }%
  \bibfield{journal}{%
  \bibinfo {journal} {J. Mater. Chem.}\ }%
  \textbf{\bibinfo {volume} {20}},\ \bibinfo {pages} {2197} (\bibinfo {year}
  {2010})%
  \bibAnnoteFile{NoStop}{B919260C}%
\bibitem{Haddon17091993}%
  \BibitemOpen
  \bibfield{author}{%
  \bibinfo {author} {\bibfnamefont{R.~C.}\ \bibnamefont{Haddon}},\ }%
  \bibfield{journal}{%
  \bibinfo {journal} {Science}\ }%
  \textbf{\bibinfo {volume} {261}},\ \bibinfo {pages} {1545} (\bibinfo {year}
  {1993})%
  \bibAnnoteFile{NoStop}{Haddon17091993}%
\bibitem{IPR}%
  \BibitemOpen
  \bibfield{author}{%
  \bibinfo {author} {\bibfnamefont{H.~W.}\ \bibnamefont{Kroto}},\ }%
  \bibfield{journal}{%
  \bibinfo {journal} {Nature}\ }%
  \textbf{\bibinfo {volume} {329}},\ \bibinfo {pages} {529} (\bibinfo {year}
  {1987})%
  \bibAnnoteFile{NoStop}{IPR}%
\bibitem{doi:10.1021/cr990332q}%
  \BibitemOpen
  \bibfield{author}{%
  \bibinfo {author} {\bibfnamefont{M.}~\bibnamefont{B\"uhl}}\ and\ \bibinfo
  {author} {\bibfnamefont{A.}~\bibnamefont{Hirsch}},\ }%
  \bibfield{journal}{%
  \bibinfo {journal} {Chemical Reviews}\ }%
  \textbf{\bibinfo {volume} {101}},\ \bibinfo {pages} {1153} (\bibinfo {year}
  {2001})%
  \bibAnnoteFile{NoStop}{doi:10.1021/cr990332q}%
\bibitem{Mostofi2008685}%
  \BibitemOpen
  \bibfield{author}{%
  \bibinfo {author} {\bibfnamefont{A.~A.}\ \bibnamefont{Mostofi}}, \bibinfo
  {author} {\bibfnamefont{J.~R.}\ \bibnamefont{Yates}}, \bibinfo {author}
  {\bibfnamefont{Y.-S.}\ \bibnamefont{Lee}}, \bibinfo {author}
  {\bibfnamefont{I.}~\bibnamefont{Souza}}, \bibinfo {author}
  {\bibfnamefont{D.}~\bibnamefont{Vanderbilt}},\ and\ \bibinfo {author}
  {\bibfnamefont{N.}~\bibnamefont{Marzari}},\ }%
  \bibfield{journal}{%
  \bibinfo {journal} {Computer Physics Communications}\ }%
  \textbf{\bibinfo {volume} {178}},\ \bibinfo {pages} {685 } (\bibinfo {year}
  {2008})%
  \bibAnnoteFile{NoStop}{Mostofi2008685}%
\bibitem{C1CP22521A}%
  \BibitemOpen
  \bibfield{author}{%
  \bibinfo {author} {\bibfnamefont{D.~E.}\ \bibnamefont{Bean}}, \bibinfo
  {author} {\bibfnamefont{J.~T.}\ \bibnamefont{Muya}}, \bibinfo {author}
  {\bibfnamefont{P.~W.}\ \bibnamefont{Fowler}}, \bibinfo {author}
  {\bibfnamefont{M.~T.}\ \bibnamefont{Nguyen}},\ and\ \bibinfo {author}
  {\bibfnamefont{A.}~\bibnamefont{Ceulemans}},\ }%
  \bibfield{journal}{%
  \bibinfo {journal} {Phys. Chem. Chem. Phys.}\ }%
  \textbf{\bibinfo {volume} {13}},\ \bibinfo {pages} {20855} (\bibinfo {year}
  {2011})%
  \bibAnnoteFile{NoStop}{C1CP22521A}%
\bibitem{doi:10.1021/jp304891e}%
  \BibitemOpen
  \bibfield{author}{%
  \bibinfo {author} {\bibfnamefont{W.-W.}\ \bibnamefont{Wang}}, \bibinfo
  {author} {\bibfnamefont{J.-S.}\ \bibnamefont{Dang}}, \bibinfo {author}
  {\bibfnamefont{J.-J.}\ \bibnamefont{Zheng}},\ and\ \bibinfo {author}
  {\bibfnamefont{X.}~\bibnamefont{Zhao}},\ }%
  \bibfield{journal}{%
  \bibinfo {journal} {The Journal of Physical Chemistry C}\ }%
  \textbf{\bibinfo {volume} {116}},\ \bibinfo {pages} {17288} (\bibinfo {year}
  {2012})%
  \bibAnnoteFile{NoStop}{doi:10.1021/jp304891e}%
\bibitem{non-IPR}%
  \BibitemOpen
  \bibfield{author}{%
  \bibinfo {author} {\bibfnamefont{A.}~\bibnamefont{Rodriguez-Fortea}},
  \bibinfo {author} {\bibfnamefont{N.}~\bibnamefont{Alegret}}, \bibinfo
  {author} {\bibfnamefont{A.~L.}\ \bibnamefont{Balch}},\ and\ \bibinfo {author}
  {\bibfnamefont{J.~M.}\ \bibnamefont{Poblet}},\ }%
  \bibfield{journal}{%
  \bibinfo {journal} {Nature Chemistry}\ }%
  \textbf{\bibinfo {volume} {2}},\ \bibinfo {pages} {955} (\bibinfo {year}
  {2010})%
  \bibAnnoteFile{NoStop}{non-IPR}%
\bibitem{genovese:014109}%
  \BibitemOpen
  \bibfield{author}{%
  \bibinfo {author} {\bibfnamefont{L.}~\bibnamefont{Genovese}}, \bibinfo
  {author} {\bibfnamefont{A.}~\bibnamefont{Neelov}}, \bibinfo {author}
  {\bibfnamefont{S.}~\bibnamefont{Goedecker}}, \bibinfo {author}
  {\bibfnamefont{T.}~\bibnamefont{Deutsch}}, \bibinfo {author}
  {\bibfnamefont{S.~A.}\ \bibnamefont{Ghasemi}}, \bibinfo {author}
  {\bibfnamefont{A.}~\bibnamefont{Willand}}, \bibinfo {author}
  {\bibfnamefont{D.}~\bibnamefont{Caliste}}, \bibinfo {author}
  {\bibfnamefont{O.}~\bibnamefont{Zilberberg}}, \bibinfo {author}
  {\bibfnamefont{M.}~\bibnamefont{Rayson}}, \bibinfo {author}
  {\bibfnamefont{A.}~\bibnamefont{Bergman}},\ and\ \bibinfo {author}
  {\bibfnamefont{R.}~\bibnamefont{Schneider}},\ }%
  \bibfield{journal}{%
  \bibinfo {journal} {The Journal of Chemical Physics}\ }%
  \textbf{\bibinfo {volume} {129}},\ \bibinfo {eid} {014109} (\bibinfo {year}
  {2008})%
  \bibAnnoteFile{NoStop}{genovese:014109}%
\bibitem{PhysRevLett.97.170201}%
  \BibitemOpen
  \bibfield{author}{%
  \bibinfo {author} {\bibfnamefont{E.}~\bibnamefont{Bitzek}}, \bibinfo {author}
  {\bibfnamefont{P.}~\bibnamefont{Koskinen}}, \bibinfo {author}
  {\bibfnamefont{F.}~\bibnamefont{G\"ahler}}, \bibinfo {author}
  {\bibfnamefont{M.}~\bibnamefont{Moseler}},\ and\ \bibinfo {author}
  {\bibfnamefont{P.}~\bibnamefont{Gumbsch}},\ }%
  \bibfield{journal}{%
  \bibinfo {journal} {Phys. Rev. Lett.}\ }%
  \textbf{\bibinfo {volume} {97}},\ \bibinfo {pages} {170201} (\bibinfo {year}
  {2006})%
  \bibAnnoteFile{NoStop}{PhysRevLett.97.170201}%
\bibitem{PhysRevB.58.3641}%
  \BibitemOpen
  \bibfield{author}{%
  \bibinfo {author} {\bibfnamefont{C.}~\bibnamefont{Hartwigsen}}, \bibinfo
  {author} {\bibfnamefont{S.}~\bibnamefont{Goedecker}},\ and\ \bibinfo {author}
  {\bibfnamefont{J.}~\bibnamefont{Hutter}},\ }%
  \bibfield{journal}{%
  \bibinfo {journal} {Phys. Rev. B}\ }%
  \textbf{\bibinfo {volume} {58}},\ \bibinfo {pages} {3641} (\bibinfo {year}
  {1998})%
  \bibAnnoteFile{NoStop}{PhysRevB.58.3641}%
\bibitem{springerlink:10.1007/s00214-005-0655-y}%
  \BibitemOpen
  \bibfield{author}{%
  \bibinfo {author} {\bibfnamefont{M.}~\bibnamefont{Krack}},\ }%
  \bibfield{journal}{%
  \bibinfo {journal} {Theoretical Chemistry Accounts: Theory, Computation, and
  Modeling (Theoretica Chimica Acta)}\ }%
  \textbf{\bibinfo {volume} {114}},\ \bibinfo {pages} {145} (\bibinfo {year}
  {2005})%
  \bibAnnoteFile{NoStop}{springerlink:10.1007/s00214-005-0655-y}%
\bibitem{PhysRevLett.77.3865}%
  \BibitemOpen
  \bibfield{author}{%
  \bibinfo {author} {\bibfnamefont{J.~P.}\ \bibnamefont{Perdew}}, \bibinfo
  {author} {\bibfnamefont{K.}~\bibnamefont{Burke}},\ and\ \bibinfo {author}
  {\bibfnamefont{M.}~\bibnamefont{Ernzerhof}},\ }%
  \bibfield{journal}{%
  \bibinfo {journal} {Phys. Rev. Lett.}\ }%
  \textbf{\bibinfo {volume} {77}},\ \bibinfo {pages} {3865} (\bibinfo {year}
  {1996})%
  \bibAnnoteFile{NoStop}{PhysRevLett.77.3865}%
\bibitem{PhysRevB.79.245401}%
  \BibitemOpen
  \bibfield{author}{%
  \bibinfo {author} {\bibfnamefont{C.~R.}\ \bibnamefont{Hsing}}, \bibinfo
  {author} {\bibfnamefont{C.~M.}\ \bibnamefont{Wei}}, \bibinfo {author}
  {\bibfnamefont{N.~D.}\ \bibnamefont{Drummond}},\ and\ \bibinfo {author}
  {\bibfnamefont{R.~J.}\ \bibnamefont{Needs}},\ }%
  \bibfield{journal}{%
  \bibinfo {journal} {Phys. Rev. B}\ }%
  \textbf{\bibinfo {volume} {79}},\ \bibinfo {pages} {245401} (\bibinfo {year}
  {2009})%
  \bibAnnoteFile{NoStop}{PhysRevB.79.245401}%
\bibitem{doi:10.1021/ar00150a005}%
  \BibitemOpen
  \bibfield{author}{%
  \bibinfo {author} {\bibfnamefont{R.~C.}\ \bibnamefont{Haddon}},\ }%
  \bibfield{journal}{%
  \bibinfo {journal} {Accounts of Chemical Research}\ }%
  \textbf{\bibinfo {volume} {21}},\ \bibinfo {pages} {243} (\bibinfo {year}
  {1988})%
  \bibAnnoteFile{NoStop}{doi:10.1021/ar00150a005}%
\bibitem{Muya2012111}%
  \BibitemOpen
  \bibfield{author}{%
  \bibinfo {author} {\bibfnamefont{J.~T.}\ \bibnamefont{Muya}}, \bibinfo
  {author} {\bibfnamefont{T.}~\bibnamefont{Sato}}, \bibinfo {author}
  {\bibfnamefont{M.~T.}\ \bibnamefont{Nguyen}},\ and\ \bibinfo {author}
  {\bibfnamefont{A.}~\bibnamefont{Ceulemans}},\ }%
  \bibfield{journal}{%
  \bibinfo {journal} {Chemical Physics Letters}\ }%
  \textbf{\bibinfo {volume} {543}},\ \bibinfo {pages} {111 } (\bibinfo {year}
  {2012})%
  \bibAnnoteFile{NoStop}{Muya2012111}%

\end{thebibliography}

\section{Acknowledgements}
We acknowledge support by the ANR through the
NANOSIM-GRAPHENE project and NEWCASTLE. This work was
performed using HPC resources from the GENCI-CINES (Grant 6194).


\newpage

\end{document}